\documentclass[12pt]{article}
\usepackage{amsmath, amssymb}
\usepackage{graphicx}
\usepackage{clara}
\usepackage[T1]{fontenc}
\usepackage{cite}

\usepackage[a4paper,margin=3cm]{geometry}

\title{Network Characteristics of Individual Pigments in Cyanobacterial Photosystem II Core Complexes}
\author{Eun Lee \and Petter Holme}
\date{Department of Energy Science, Sungkyunkwan University, Suwon 440-746, Korea}

\begin{document}
\maketitle

\begin{abstract}
Part of the excitation energy transfer (EET) characteristics of the photosystem II (PSII) comes from the interconnection between pigments. To understand the correlation between the EET and the pigments' interaction structure, we construct a network from the EET rates, which are related to both the distance between the pigments (chlorophylls and pheophytins) and their spatial orientations. Especially, we investigate how well the PS II core complex's EET functionality can be explained by using only the network topology in Thermosynechococcus vulcanus 1.9 \AA. Starting from the F\"orster theory, we construct a network of EET pathways. For an analysis of the network structure, we calculate common network-structural measures like betweenness centrality, eigenvector centrality and weighted clustering. These measures can reflect the role of individual pigments in the EET network. In our work, we found that some well-known properties were reproduced by the network analysis of the simplified network, which means that the topology of the network encodes functionally relevant information. For example, from the network structural analysis, we can infer that most of the chlorophyll molecules (chlorophylls) in the pigment-protein complex CP47 have a heightened probability of transferring energy compared with other chlorophylls. We also see that the active branch chlorophylls in the reaction center are characterized by a high eigenvector centrality, a high betweenness centrality, and a low weighted clustering coefficient. This is indicative of functionally important vertices.
\end{abstract}

\section{Background}
Photosynthetic organisms are able to generate chemical energy from the sun. The efficiency of harvesting solar energy is almost unity. This impressive mechanism begins from the excitation energy transfer (EET) in Photosystem II (PSII), which is located on thylakoid membranes of chloroplasts in photosynthetic cells \cite{ref1}. Especially, many experiments have focused on relating the structure and EET properties \cite{ref2,ref3,ref4}. Because chlorophylls in these pigment-protein complexes are arranged in a peculiar pattern, many researchers have studied the overall EET of the system by the orientations and distances between the pigments in the protein environment \cite{ref5,ref6}. In the case of the PSII core complex, this kind of research is possible since the crystal structure has been mapped out at resolutions from 1.9 \AA{} to 3.8 \AA{} in two cyanobacteria, \textit{Thermosynechococcus elongatus} \cite{ref7,ref8,ref9} and \textit{Thermosynechococcus vulcanus} \cite{ref10,ref11}. In the PSII core complex crystal structure, there are 70 chlorophylls containing the so-called D1, D2 protein complex and CP47, CP43 complex. To understand the mechanisms of the PSII EET exactly, we have two options. One is adding additional variables estimated by spectroscopic studies. The other is simplifying the EET system information to understand the big picture. Some research has assumed the EET dynamics to be an interactive network to improve the general understanding of Photosystem I (PSI) EET \cite{ref12,ref13}. The studies are applied to PSI and focus on the theoretical calculation. In our research, we aim to see the big picture of EET in PSII by modeling the photosynthetic EET system as a network of nodes and links: pigments (chlorophyll and pheophytin) are treated as single nodes, and links connect nodes with EET rates high enough. In this work, we investigate the hypothesis that each pigment (chlorophylls and pheophytins) in the PSII core complex can be viewed as a node in a network, and that the network topology explains some of the functionality of the EET.

\section{Methods}
In large interconnected systems, the way pigments are connected---the network topology---can explain (several features of) the functionality of the system and its constituents. One of the most important ways to characterize a node's role in a network is by measuring its centrality \cite{ref14,ref15}. In fact, there are a number of centrality measures, each corresponding to one aspect of the umbrella concept of centrality. In this section, we will present the calculation and algorithm for constructing the simplified network, the centrality measures, and several parameters related to other structures of the network that we use.

\subsection{Calculation of the Excitation-Energy Transfer (EET) Rate}
As mentioned above, we use the dimeric PSII core complex. The direction of the dipole-moment vector of the lowest excited ($Q_{y}$) state of the chlorophyll and the pheophytin approximately follows a line connecting the NB and ND atoms in the porphyrin ring of the chlorophyll and pheophytin \cite{ref16,ref17}. We obtain the orientation of the dipole vector from the crystal structure. The EET rate can represent the probability of energy transfer between pairs of nodes. We will use the EET rate as the transfer probability, which could indirectly represent the amount of energy transferred through a node $x$ per unit time.

To calculate the EET rate, we should first determine the coupling strength. $W_{ij}$ denotes the coupling strength between the ith and the jth chlorophylls. The electric coupling $W_{ij}$ can be divided into two terms
\begin{equation}
W_{ij}=W_{ij}^{c}+W_{ij}^{ex}
\end{equation}
where $W_{ij}^{c}$ corresponds to a direct coulomb term and $W_{ij}^{ex}$ corresponds to an electron exchange term. In the PSII core complex, all of the distances between Mg atoms of pigments are mostly greater than 7 \AA{} so we assume that the Coulomb interaction dominates the electric coupling strength \cite{ref12}. From this assumption, it follows that exciton states are fairly well localized to individual pigments, which is a basic assumption of F\"orster theory \cite{ref18}. Therefore, we keep only the direct Coulomb term $W_{ij}^{c}$. For the calculation of the Coulomb term, we assume the dipole-dipole interaction between transition dipole moments of the chlorophyll $Q_{y}$ states. With this approximation, the couplings are given by
\begin{equation}
W_{ij}=\alpha\left(\frac{d_{i}\cdot d_{j}}{r_{ij}^{3}}-\frac{3(r_{ij}\cdot d_{i})(r_{ij}\cdot d_{j})}{r_{ij}^{5}}\right)
\end{equation}
where $d_{i}$ are unit vectors along the transition dipole moments from the ground state to the $Q_{y}$ state of the $i$'th chlorophyll and $r_{ij}$ is the vector connecting the Mg atoms of chlorophyll i and j. Here, we use the value of the parameter $\alpha$ as 116,000 \AA$^{3}\text{cm}^{-1}$ \cite{ref12}.

In addition to the coupling strength between nodes, the energy of nodes can play an important role in determining the direction of excitation energy flow. Because the nodes correspond to chlorophylls, we can measure the energy of nodes from the absorption spectrum of chlorophyll. We used the site energy of special-pair chlorophylls as $14,630\text{ cm}^{-1}$ and the others are $14,814\text{ cm}^{-1}$ \cite{ref19}.

After the setting of the site energy and the coupling strength, we calculate the EET rate between chlorophylls based on F\"orster theory \cite{ref13} by using the appropriate equation. The EET rate from pigment i to pigment j is calculated by
\begin{equation}
T_{ij}=\frac{4\pi^{2}}{h}|W_{ij}|^{2}J_{ij}
\end{equation}
Here, $J$ represents the spectral overlap of the donor emission and acceptor absorption. We can calculate this quantity as follows:
\begin{equation}
J_{ij}=\int S_{i}^{D}(E)S_{j}^{A}(E)dE
\end{equation}
In this formula, $S^{D}_{i}(E)$ is the donor's emission spectrum, and $S^{A}_{j}(E)$ is the acceptor's absorption spectrum. We approximate these spectra as Gaussians as follows:
\begin{equation}
P_{DA_{x}}=\frac{r_{DA_{x}}}{\sum_{n=1}^{N}r_{DA_{n}}}
\end{equation}
where $r_{DA_{x}}$ means the transfer rate from donors to acceptor.
\begin{subequations}
\begin{equation}
S_{i}^{D}(E)=\frac{1}{\sqrt{2\pi}\nu}\exp\left[-\frac{(E_{i}-S-E)^{2}}{2\nu}\right]
\end{equation}
\begin{equation}
S_{j}^{A}(E)=\frac{1}{\sqrt{2\pi}\nu}\exp\left[-\frac{(E_{j}-E)^{2}}{2\nu}\right]
\end{equation}
\end{subequations}
where $E_{i}$ and $E_{j}$ are the absorption peaks for the pigments and $S$ represents the Stokes shift. We take the Stokes shift to be equal to $180\text{ cm}^{-1}$ at room temperature and assume the widths for the emission and the absorption spectra to be the same value \cite{ref19,ref20}. Combining the coupling strength and spectral overlap, we are able to get the EET rate between pigments of PSII dimeric core complex.

\subsection{Constructing the Chlorophyll Network: Overview}
To construct the EET network, we start from the crystal structure of the PSII core complex of T. vulcanus obtained with 1.9 \AA{} resolution from the Protein Data Bank \cite{ref11}. This database contains information about the coordinates of the Mg atoms in chlorophylls and their orientations. From this information, one can assign an EET rate (explained above) to pairs of chlorophylls. Using the minimization algorithm, we make a simplified network that keeps all connections. We assign links between nodes to minimize the number of links while maintaining the network's connectivity. Below, we describe the simplified network.

\subsection{Details of the Network Construction and Dynamic Simulation}
Based on the calculated EET rate, we construct the networks of the PSII core complex as described above. Starting from a fully connected network, we delete links in order of increasing weight. We stop this deletion process before deleting the link that would disconnect the network. The cut-off weight for this process is $4.2\times10^{10}\text{ s}^{-1}$. Once constructed, the network has 74 nodes and 278 links. Then we use the simplified network to distribute the EET using the algorithm described below. In real spectroscopic experiments, exclusively exciting a pigment is almost impossible, which is what we do in our simulations to find the functionality of pigments in excitation transmission. We excite every node (chlorophyll and pheophytin) and monitor the ingoing and the outgoing flows of excitation energy through each node until no excitation is left.

\begin{figure}
\centering\includegraphics[width=0.3\linewidth]{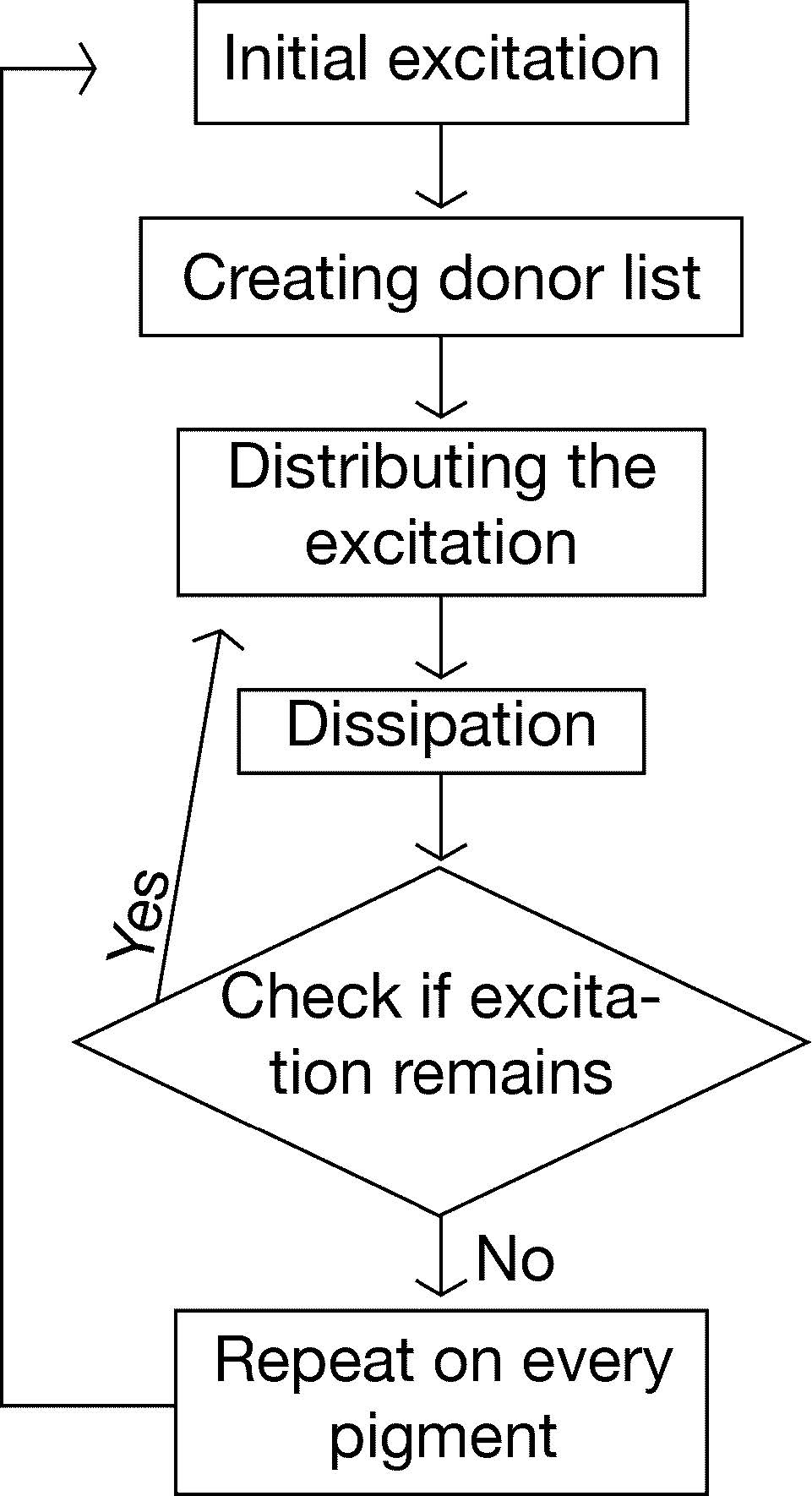}
\caption{Flow chart for the EET distribution algorithm on
an EET network.}
\label{fig:1}
\end{figure}

Our simulation algorithm is illustrated in Fig~\ref{fig:1}. First, we apply one unit of excitation to a single node. Second, we transfer the excitation to neighbors of the excited node with a probability proportional to the EET rate. Third, in this process, we choose one of two ways for the excitation to disappear in the network. One is a self-dissipation pathway for every chlorophyll, in which one unit of excitation leaves the network in one nanosecond \cite{ref21}. The other one, for nodes in the reaction center, is charge separation, which occurs on picosecond timescales \cite{ref22}. To solve the technical problem, we add auxiliary, dummy nodes to express the effects of charge separation and self-dissipation. For these two cases, we set the weights of additional nodes to the self-dissipation rate ($10^{9}$) and the charge separation rate ($10^{12}$). The active-branch chlorophyll in the reaction center (labeled 1604, 2604) and the accessory chlorophyll (labeled 1606, 2606) in each reaction center's active branch are controversial regarding charge separation \cite{ref23,ref24}. In this work, we assume that charge separation will occur on the special pair of active chlorophyll. Calculating the excitation population by using a master equation would be similar to this algorithm. We construct this, however, because we want to measure not only the final population but also the total sum of in- and out-flows in each node.

\subsection{Degree Centrality}
The simplest measure of network centrality we measure is called degree centrality \cite{ref14,ref15,ref25}. Degree centrality is simply the number of other nodes to which a node is connected. As opposed to betweenness and eigenvector centralities (discussed below), degree centrality is local in the sense that the degree centrality of $i$ is independent of the network outside of $i$'s neighborhood. If a network property is determined entirely by a local quantity, parts far away (involving many nodes connected indirectly by paths of links) are not important for the dynamics of the system.

\subsection{Betweenness Centrality}
Another centrality concept comes from thinking about the flow of a dynamic system through a node. If we assume that the flow originates between pairs of nodes at equal rates and travels through the network along the shortest paths, then the amount of traffic through a node would be proportional to its betweenness centrality \cite{ref25,ref14,ref15}. More specifically, let $\sigma(i,j)$ be the number of shortest paths between $i$ and $j$ (a shortest path does not have to be unique), and let $\sigma_{l}(i,j)$ be the number of shortest paths between $i$ and $j$ that pass $l$. Then $l$'s betweenness centrality is
\begin{equation}
C_{B}(l)=\sum_{j}\sum_{i\ne j}\frac{\sigma_{l}(i,j)}{\sigma(i,j)}
\end{equation}
This definition holds for both weighted and unweighted networks, although for weighted networks the shortest path is more often unique and so the denominator is usually strictly one.

\subsection{Eigenvector Centrality}
A third concept of centrality comes from the observation that if the neighbors of a node are central, then this node is probably central, too. We denote this centrality as $C_{E}(i)$ and assume that it can be transferred between connected nodes. If we let the superscript denote the time of the spreading, then one can summarize the above concept as
\begin{equation}
C_{E}^{t+1}(i)=\lambda^{-1}\sum_{j\in\Gamma_{i}}C_{E}^{t}(j)
\end{equation}
Where $\Gamma_{i}$ is the neighborhood of $i$ (i.e., all vertices that have a link to $i$), and $\lambda$ is a constant to ensure that the iterations to converge. We are interested in the long-time limit when $t\rightarrow\infty$, so we can set $t=t+1$. This means that the $C_{E}$ can be phrased as an eigenvector problem:
\begin{equation}
C_{E}=\lambda^{-1}AC_{E}
\end{equation}
where $A$ is the adjacency matrix (where the $ij$ element is 1 if there is a link between $i$ and $j$ and 0 otherwise) and $C_{E}$ is the vector of centrality values for all vertices.

\subsection{Clustering Coefficient}
There are types of network structure other than centrality. The clustering coefficient \cite{ref14,ref15} is a measure of how well one member of relatively many triangles (relative to the maximal number given the number of neighbors). Triangles introduce redundancy in the sense that paths that include a given link become only one step longer if the link vanishes. We define the clustering coefficient of node $i$ as
\begin{equation}
c(i)=\frac{t_{i}}{\binom{k_{i}}{2}}
\end{equation}
Where $t_{i}$ is the number of triangles that $i$ is a part of. The binomial factor $\binom{k_{i}}{2}=\frac{1}{2}k_{i}(k_{i}-1)$ gives the maximal number of triangles given the degree of $i$.

\begin{figure}
\centering\includegraphics[width=0.6\linewidth]{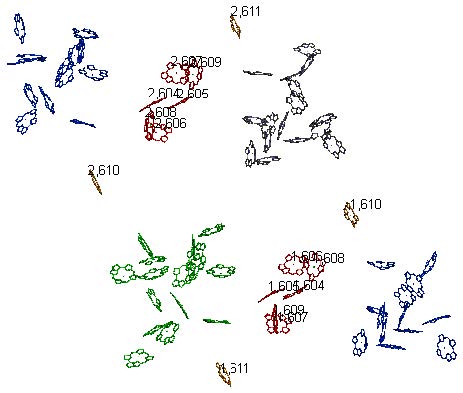}
\caption{PSII of the \textit{Thermosynechococcus
vulcanus}. The figure shows the crystal structure in 1.9 \AA resolution (3ARC in Protein Data Bank) of pigments (chlorophylls and pheophytins). For convenience, we separated the dimeric structure into two monomeric structures by adding the prefixes `1' and `2'. Deep blue color shows reaction center components and linker chlorophylls which are 1610, 1611, 2610 and 2611. (Figure produced with Discovery Studio Visualizer.)}
\label{fig:2}
\end{figure}

\begin{figure}
\centering\includegraphics[width=0.6\linewidth]{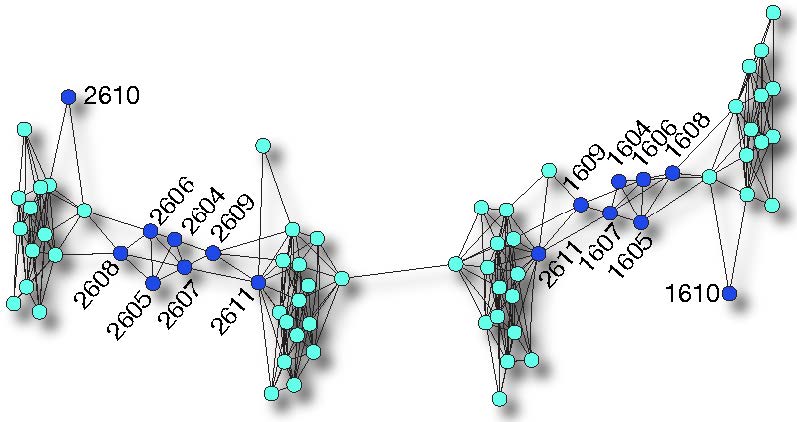}
\caption{The EET network of PS II. We used
and modified the chlorophyll numbering of PDB entry 3ARC~\cite{ref11}. Node numbers 1604, 1605, 2604, and 2605 represent special pairs of reaction centers. This graph seeks to display the network topology in a clear way. Two nodes are close if they have a short network distance between one another (the position is only indirectly related to the spatial configuration
of the chlorophylls).}
\label{fig:3}
\end{figure}

\section{Results}
\subsection{Basic Properties of the Network and Dynamics}
First, we discuss the basic structure of the excitation networks derived from the relative locations and orientations of the pigments. In Fig.~\ref{fig:2}, we display the crystal structure of chlorophylls and pheophytins. The EET network is shown in Fig.~\ref{fig:3}. The layout of the EET network
in Fig.~\ref{fig:3} is not the same as the pigment locations in the crystal structure (Fig.~\ref{fig:2}).
A short distance in this plot represents a short graph distance, which is correlated with, but not identical to, the physical distance. The EET network has 70 chlorophylls and 4 pheophytins as nodes, with 278 links between chlorophylls based on EET rates. The size of the network is rather small compared with other typical empirical datasets studied in the research field of complex networks \cite{ref14}. For simplicity, we assumed the links were bidirectional. This is possible because we need the averaged EET rate between the nodes. To quantify the importance of a node in excitation-transfer dynamics, we propose measuring the total passed excitation (TPE). We calculated the TPE by summing the ingoing and the outgoing excitation energies through each node. This shows how much excitation flows in and out of the node.

\begin{figure}
\centering\includegraphics[width=0.6\linewidth]{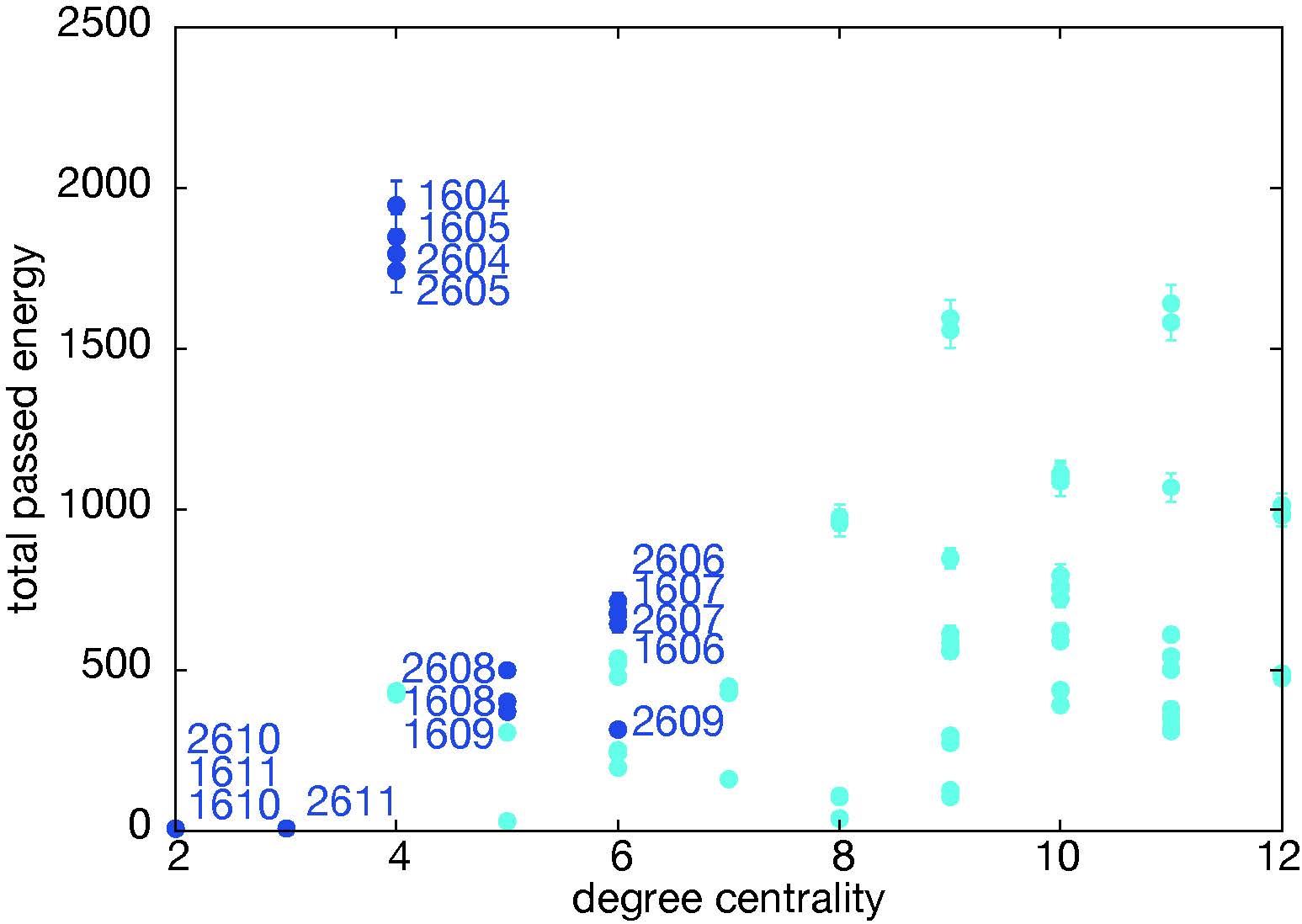}
\caption{Scatter plot of the total passed energy versus the degree centrality. The nodes of the reaction center are plotted with a darker shade. Error bars correspond to one standard error (the standard deviation of the mean).}
\label{fig:4}
\end{figure}

\subsection{Degree Centrality}
In the case of degree centrality \cite{ref14,ref15}, most of the nodes show a positive correlation with the TPE value (Fig.~\ref{fig:4}). The special pairs display a different characteristic, where the TPE is exceptionally large, but the degree centrality is only intermediate. This means that even though these nodes (special pair chlorophylls) are connected to only a limited number of others, a large amount of energy passes through them. Focusing on the main cloud of points, 1613, 1614, 2613, and 2614 (all in the CP47 complex) are somewhat outliers, having both very large degrees and TPE. Furthermore, we note that within the special pairs, the chlorophylls belonging to the active branch of the reaction center have higher degree centralities than the ones in the inactive branch.

\begin{figure}
\centering\includegraphics[width=0.6\linewidth]{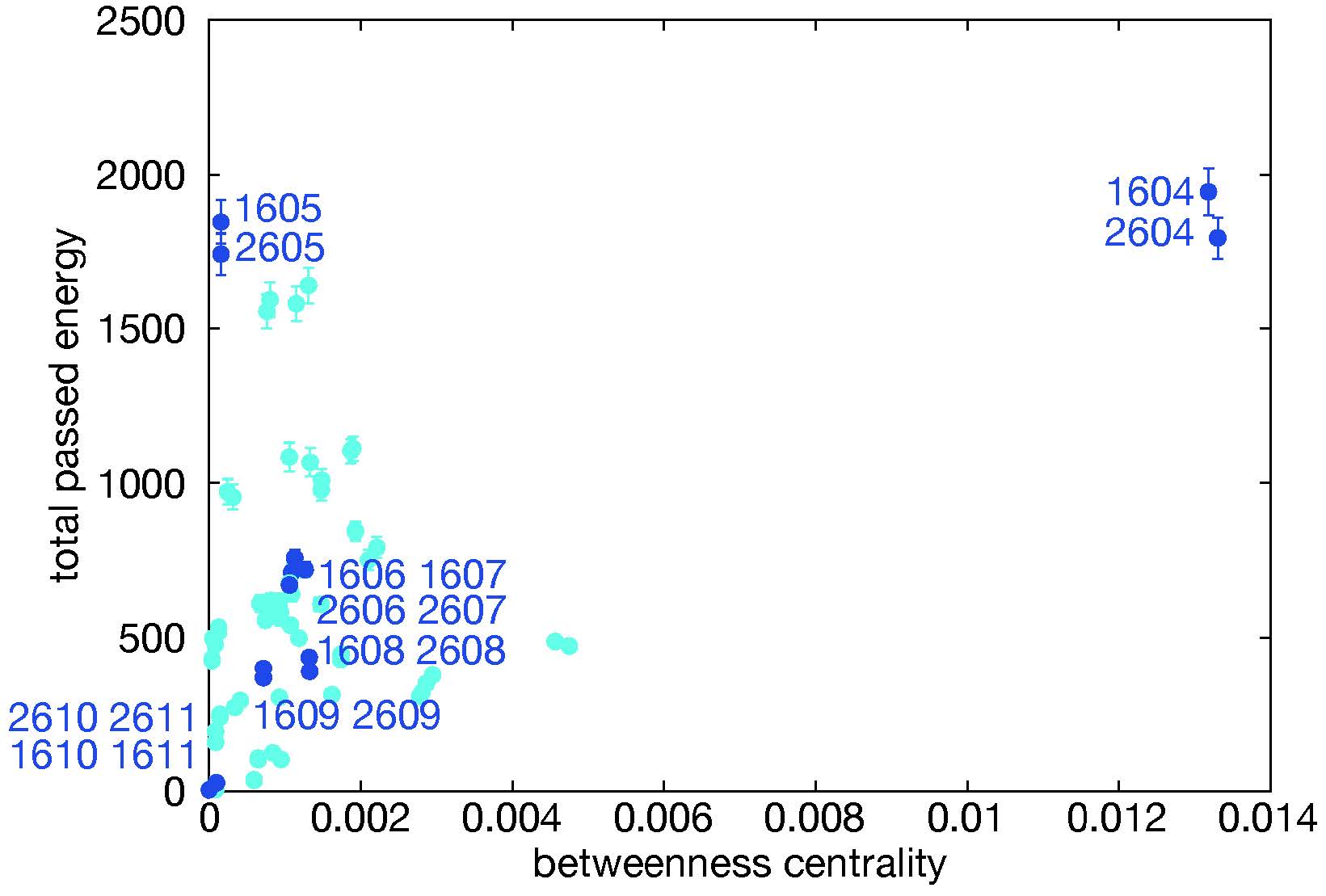}
\caption{Scatter plot of the total passed energy
versus the eigenvector centrality. Symbols are the same
as in Fig.~4.}
\label{fig:5}
\end{figure}

\subsection{Eigenvector Centrality}
Results for eigenvector centrality can be seen in Fig.~\ref{fig:5}. Many of the active branch's chlorophylls are ranked higher than inactive branch molecules, which could indicate that the active branch has a more important role in the EET. On the other hand, because the TPE is the quantity of primary functional importance, the main conclusion is that eigenvector centrality cannot capture it. Eigenvector centrality is, in a sense, an extension of degree centrality to a global quantity. Because degree centrality is more strongly correlated with TPE, charge transfer is a rather local process in this model, with shorter excitation cascades.

\begin{figure}
\centering\includegraphics[width=0.6\linewidth]{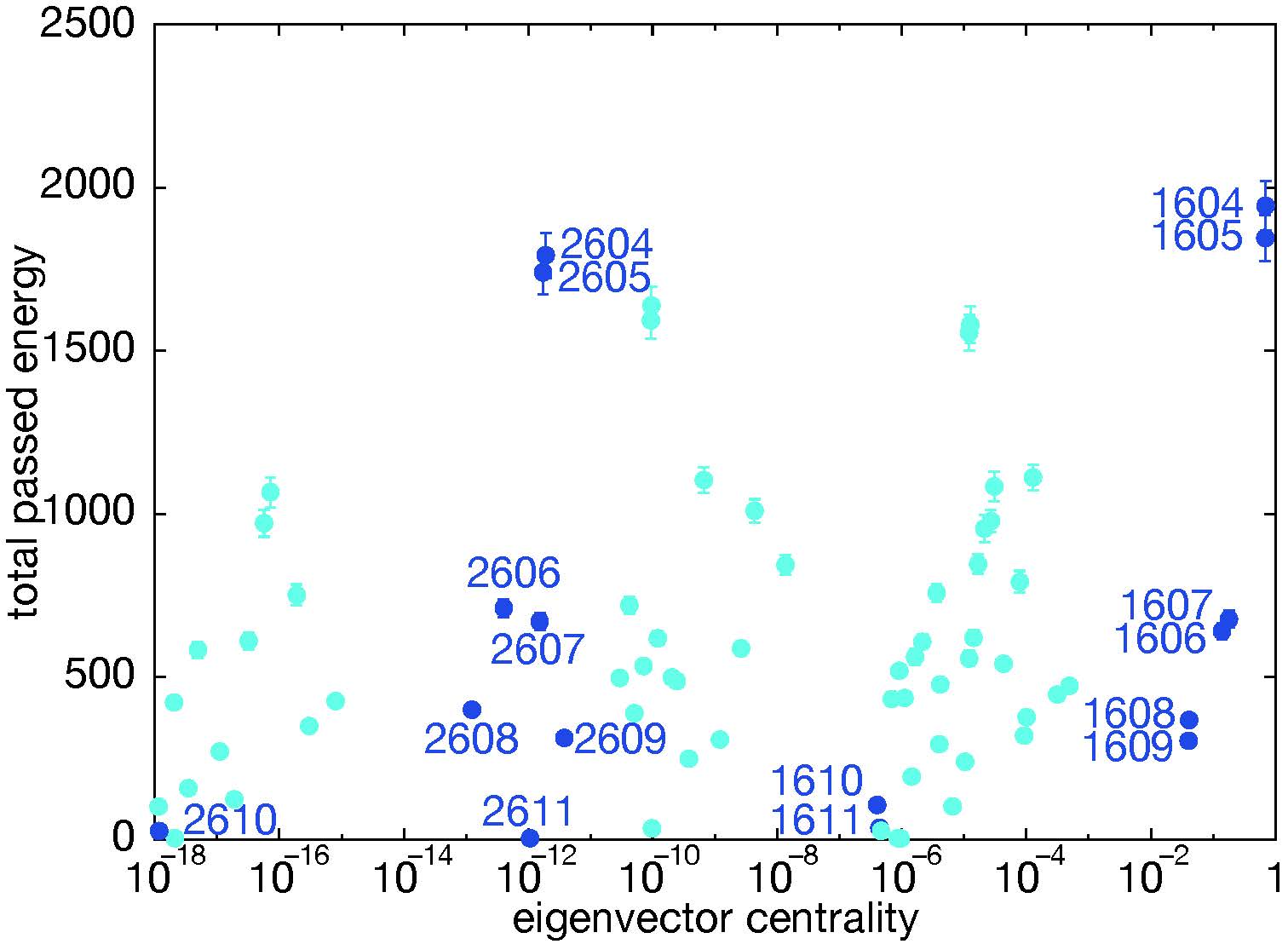}
\caption{Scatter plot of the total passed energy
versus the betweenness centrality. The nodes of the reaction center are plotted with a darker shade. Error bars correspond to one standard error.}
\label{fig:6}
\end{figure}

\subsection{Betweenness Centrality}
Looking at the betweenness centrality \cite{ref14,ref15} in Fig.~\ref{fig:6}, we see a difference between the active and inactive branches of the reaction center even more clearly than for the degree centrality. The special pair of the active branch has an almost 14 times higher betweenness centrality than the special pair of the inactive branch. The nodes that are close in the network can still have different betweenness values (which is typical for this measure). The reason is that some nodes, although central in other respects, can lie a bit off the main avenues of the shortest paths and, thus, have many fewer shortest paths passing through them than their neighbors. We note that the motivation of betweenness centrality assumes objects traveling on the network that actively try to take the shortest paths. This is, of course, not the case in the system we study; nevertheless, betweenness highlights the difference between the active and inactive branches of the reaction center. Moreover, the network structure can be optimized so that the excitation with a heightened probability travels short paths to the reaction center.

\begin{figure}
\centering\includegraphics[width=0.6\linewidth]{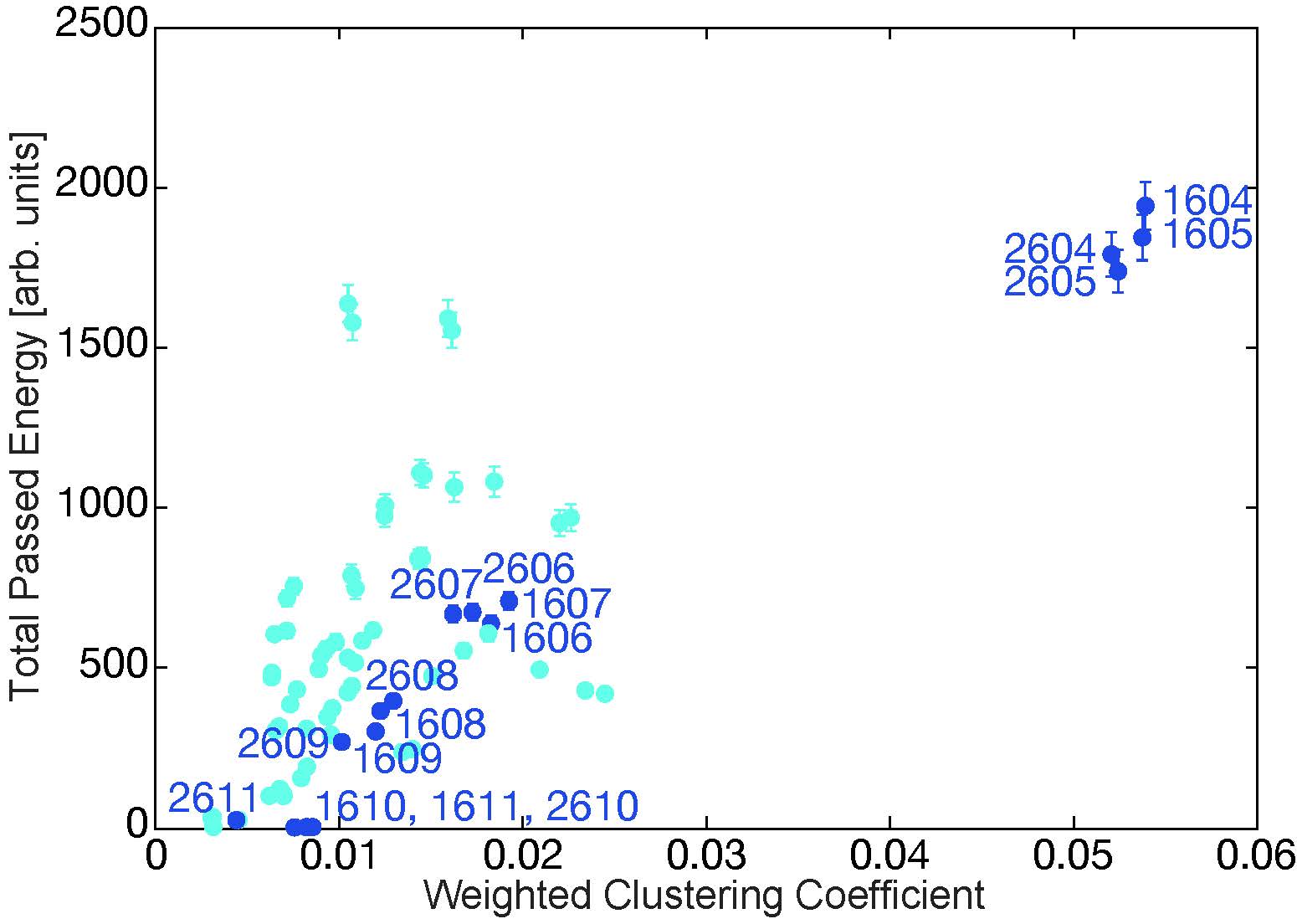}
\caption{Scatter plot of the total passed energy versus the weighted clustering coefficient. Symbols are the same as in Fig.~4.}
\label{fig:7}
\end{figure}

\subsection{Clustering Coefficient}
The weighted clustering coefficient measures how connected the neighborhood of a node is \cite{ref27}. A large value of this quantity means that charge is likely to pass through the neighborhood of a node rather than through the node itself. Nothing a priori dictates a positive correlation between the weighted clustering coefficient and the TPE; indeed, many networks show a negative correlation between the clustering coefficient and degree \cite{ref28} (Fig.~\ref{fig:7}). In this case, no clear outliers exist; rather, a positively correlated point cloud exists. The special pairs are the nodes with the highest weighted clustering and the highest TPE.

\section{Discussion}
In our investigation of the relation between the network topology and the EET dynamics of PSII, we found that network structural measures that were designed to capture some aspects of a pigment's importance were positively correlated with a measure of the functional importance. This means that much of the dynamics of the PSII EET can qualitatively be explained by the network topology. One such example is the active-branch chlorophyll that is an outlier in many of our correlation plots; in other words, it seems to be more important in EET because of its position in the network. The chlorophyll in the active branch shows higher betweenness and weighted clustering coefficient than the chlorophylls in the inactive branch. This is especially true for the betweenness centralities of the special pair that has almost 14 times higher values in the active than in the inactive branch. From these results, we can infer that a chlorophyll in the active branch is located in such a way as to absorb the excitation better than a chlorophyll in the inactive branch.

Even though we estimate the EET before the charge separation process, the chlorophyll in the active branch shows high functionality. In the literature, quite a consensus exists that the chlorophylls in the active branch of the special pair are important for charge separation \cite{ref24,ref29}. Combining those results and ours, we can infer that the network characteristics, which are represented as the probability of getting EET, may be correlated to charge separation.

Chlorophylls 1613, 1614, 2613, and 2614 form another group with conspicuous structural features. These chlorophylls are included in the CP47 complex and are peculiarly central in the network given their TPE values. Furthermore, a second monomer has a higher eigenvector centrality than the first. From the construction of the eigenvector centrality, this suggests that the second monomer could have a somewhat more important role in the relaxation process after photon absorption. A direct experimental verification of this conclusion is difficult, but it is our hypothesis for future studies. One more future topic would be that even though the dimer structure makes the system resilient to failure by introducing redundancy, evolution has favored one branch.

On the other hand, the eigenvector centrality is rather weakly correlated with the dynamic measure of importance, TPE. Our hypothesis is that the TPE contains functionally more relevant information. However, once again, how to weigh this different information will ultimately need experimental evaluation.

To summarize, we see that network theory confirms well-known structural properties of the bacterial PS II core complex and verifies their functions in EET with a broad picture. This can be useful in future studies to make a sketchy overview of systemic organization and to map out the optimized structure of photosynthesis. In future work, we also hope to extend our assumptions behind the network construction, such as the F\"orster theory, and to study the efficiency of the network as a whole, not just the individual nodes. We anticipate future investigations of photosynthesis and similar spatially-constrained complex biochemical processes by using network theory.

\bibliographystyle{abbrv}
\bibliography{bib}

@book{ref1, 
    author={R. E. Blankenship}, 
    title={Molecular mechanisms of photosynthesis}, 
    publisher={Blackwell Science}, 
    address={Hoboken NJ}, 
    year={1999}
}

@article{ref2, 
    author={N. P. Paw{\l}owicz and M. L. Groot and I. H. M. van Stokkum and J. Breton and R. van Grondelle}, 
    title={Charge separation and energy transfer in the photosystem II core complex studied by femtosecond midinfrared spectroscopy},
    journal={Biophys. J.}, 
    volume={93}, 
    pages={2732}, 
    year={2007}
}

@article{ref3, 
    author={G. Raszewski and T. Renger}, 
    title={Light harvesting in photosystem II core complexes is limited by the transfer to the trap: can the core complex turn into a photoprotective mode?},
    journal={J. Am. Chem. Soc.}, 
    volume={130}, 
    pages={4431}, 
    year={2008}
}

@article{ref4, 
    author={V. I. Novoderezhkin and E. Romero and J. P. Dekker and R. van Grondelle}, 
    title={Excitation dynamics in intact core complexes of photosystem II},
    journal={Chem. Phys. Chem.}, 
    volume={12}, 
    pages={691}, 
    year={2011}
}

@article{ref5, 
    author={J. Adolphs and F. M\"uh and M.-A. Madjet and M. S. am Busch and T. Renger}, 
    title={Structure-based calculations of optical spectra of photosystem I suggest an asymmetric light-harvesting process},
    journal={J. Am. Chem. Soc.}, 
    volume={132}, 
    pages={3331}, 
    year={2010}
}

@article{ref6, 
    author={M. Byrdin and P. Jordan and N. Krauss and P. Fromme and D. Stehlik and E. Schlodder}, 
    title={Light harvesting in photosystem I: modeling based on the 2.5-{\AA} structure of photosystem I from Synechococcus elongatus},
    journal={Biophys. J.}, 
    volume={83}, 
    pages={433}, 
    year={2002}
}

@article{ref7, 
    author={A. Zouni and others}, 
    title={Crystal structure of photosystem II from Synechococcus elongatus at 3.8 {\AA} resolution},
    journal={Nature}, 
    volume={409}, 
    pages={739}, 
    year={2001}
}

@article{ref8, 
    author={K. N. Ferreira and T. M. Iverson and K. Maghlaoui and J. Barber and S. Iwata}, 
    title={Architecture of the photosynthetic oxygen-evolving center},
    journal={Science}, 
    volume={303}, 
    pages={1831}, 
    year={2004}
}

@article{ref9, 
    author={A. Guskov and others}, 
    title={Cyanobacterial photosystem II at 2.9-{\AA} resolution and the role of quinones, lipids, channels and chloride},
    journal={Nature Struct. Mol. Biol.}, 
    volume={16}, 
    pages={334}, 
    year={2009}
}

@article{ref10, 
    author={N. Kamiya and J. R. Shen}, 
    title={Crystal structure of oxygen-evolving photosystem II from Thermosynechococcus vulcanus at 3.7-{\AA} resolution},
    journal={Proc. Natl. Acad. Sci. USA}, 
    volume={100}, 
    pages={98}, 
    year={2003}
}

@article{ref11, 
    author={Y. Umena and others}, 
    title={Crystal structure of oxygen-evolving photosystem II at a resolution of 1.9 {\AA}},
    journal={Nature}, 
    volume={473}, 
    pages={55}, 
    year={2011}
}

@article{ref12, 
    author={M. K. Sener and others}, 
    title={Excitation migration in trimeric cyanobacterial photosystem I},
    journal={J. Phys. Chem. B}, 
    volume={106}, 
    pages={7948}, 
    year={2002}
}

@article{ref13, 
    author={M. K. Sener and C. Jolley and A. Ben-Shem and P. Fromme and N. Nelson and R. Croce and K. Schulten}, 
    title={Evolution of the excitation transfer network in photosystem I from cyanobacteria to plants},
    journal={Biophys. J.}, 
    volume={89}, 
    pages={1630}, 
    year={2005}
}

@book{ref14, 
    author={M. E. J. Newman}, 
    title={Networks: An Introduction}, 
    publisher={Oxford University Press}, 
    address={Oxford}, 
    year={2010}
}

@book{ref15, 
    author={E. Estrada}, 
    title={The structure of complex networks: Theory and applications}, 
    publisher={Oxford University Press}, 
    address={Oxford}, 
    year={2011}
}

@article{ref16, 
    author={T. Renger and R. A. Marcus}, 
    title={On the relation of protein structural dynamics and exciton relaxation in pigment-protein complexes: An estimation of the spectral density and a theory for the calculation of optical spectra},
    journal={J. Phys. Chem. B}, 
    volume={106}, 
    pages={1809}, 
    year={2002}
}

@book{ref17, 
    author={H. Van Amerongen and L. Valkunas and R. van Grondelle}, 
    title={Photosynthetic Excitons}, 
    publisher={World Scientific}, 
    address={Singapore}, 
    year={2000}, 
    note={p. 77}
}

@article{ref18, 
    author={Y. Mino and G. R. Fleming}, 
    title={Isotropic, anisotropic, and single-molecule spectra of the light-harvesting complex II},
    journal={Chem. Phys.}, 
    volume={275}, 
    pages={355}, 
    year={2002}
}

@article{ref19, 
    author={G. Raszewski and W. Saenger and T. Renger}, 
    title={Theory of optical spectra of photosystem II core complexes: evidence for a red chlorophyll a pigment},
    journal={Biophys. J.}, 
    volume={88}, 
    pages={9868}, 
    year={2005}
}

@article{ref20, 
    author={G. Raszewski and T. Renger}, 
    title={Light harvesting in photosystem II core complexes is limited by the transfer to the trap: can the core complex turn into a photoprotective mode?},
    journal={J. Am. Chem. Soc.}, 
    volume={130}, 
    pages={4431}, 
    year={2008}
}

@article{ref21, 
    author={Y. Miloslavina and M. Szczepaniak and G. M\"uller and J. Sander and M. Nowaczyk and M. R\"ogner and A. R. Holzwarth}, 
    title={Charge separation kinetics in intact photosystem II core complexes is trap-limited. A picosecond fluorescence study},
    journal={Biochem.}, 
    volume={45}, 
    pages={2436}, 
    year={2006}
}

@article{ref22, 
    author={N. P. Paw{\l}owicz and M. L. Groot and I. H. M. van Stokkum and J. Breton and R. van Grondelle}, 
    title={Charge separation and energy transfer in the photosystem II core complex studied by femtosecond midinfrared spectroscopy},
    journal={Biophys. J.}, 
    volume={93}, 
    pages={2732}, 
    year={2007}
}

@article{ref23, 
    author={T. Cardona and A. Sedoud and C. Nicholas and A. W. Rutherford}, 
    title={Charge separation in photosystem II: A comparative and evolutionary overview},
    journal={Biochim. Biophys. Acta.}, 
    volume={1817}, 
    pages={26}, 
    year={2012}
}

@article{ref24, 
    author={I. V. Shelaev and others}, 
    title={Ultrafast primary photoprocesses in photosystem II core complexes},
    journal={J. Photoch. Photobio. B}, 
    volume={104}, 
    pages={44}, 
    year={2011}
}

@article{ref25, 
    author={G. Sabidussi}, 
    title={The centrality index of a graph},
    journal={Psychometrika}, 
    volume={31}, 
    pages={581}, 
    year={1966}
}

@article{ref27, 
    author={J. Saram\"aki and M. Kivel\"a and J. P. Onnela and K. Kaski and J. Kert\'esz}, 
    title={Generalizations of the clustering coefficient to weighted complex networks},
    journal={Phys. Rev. E}, 
    volume={75}, 
    pages={027105}, 
    year={2007}
}

@article{ref28, 
    author={E. Ravasz and A. L. Somera and D. A. Mongru and Z. N. Oltvai and A.-L. Barab\'asi}, 
    title={Hierarchical organization of modularity in metabolic networks},
    journal={Science}, 
    volume={297}, 
    pages={1551}, 
    year={2002}
}

@article{ref29, 
    author={L. Huang and N. Ponomarenko and G. P. Wiederrecht and D. M. Tiede}, 
    title={Protein environmental modulation of the exciton states and energy transfer in the photosynthetic complex},
    journal={Proc. Natl. Acad. Sci. USA}, 
    volume={109}, 
    pages={4851}, 
    year={2012}
}

\end{document}